\documentclass[twoside,slac_one]{revtex4}
\usepackage{graphicx}
\usepackage{fancyhdr}
\usepackage{amsmath} 
\usepackage{bm}
\usepackage{amsxtra}
\usepackage{amssymb}
\usepackage{amsthm}
\usepackage{latexsym}
\usepackage{lscape}

\begin{document}

\title{Relating $\bf{B_s}$ Mixing and $\bm{B_s \to \mu^+\mu^-}$ 
with New Physics -- An Update}

%

\author{E. Golowich} 
\affiliation{Physics Department, University of Massachusetts, 
Amherst, MA, USA}
%
%

\begin{abstract}
This document describes my talk (based on work by 
JoAnne Hewett, Sandip Pakvasa, Alexey Petrov, 
Gagik Yeghiyan and myself) given 
at the 2011 Meeting of the 
Division of Particles and Fields of the American Physical 
Society (8/9/11-8/13/11) hosted 
by the Physics Department at Brown University.
We perform a study of the Standard Model (SM) fit to 
the mixing quantity $\Delta M_{B_s}$ in order to bound 
contributions of New Physics to $B_s$ mixing.  
We then use this to explore the branching fraction of 
$B_{s} \to \mu^+\mu^-$ in several models of New Physics (NP).  
In some cases, this constrains NP amplitudes for 
$B_{s} \to \mu^+\mu^-$ to lie below the SM component. 
\end{abstract}

\maketitle

\thispagestyle{fancy}


\section{Introduction}
Here, I describe a calculation~\cite{Golowich:2011cx}  
carried out by JoAnne Hewett, Sandip 
Pakvasa, Alexey Petrov, Gagik Yeghiyan and 
myself (hereafter denoted collectively as GHPPY) in which 
we study the transition ${B_s \to \mu^+\mu^-}$ in the context 
of both the Standard Model (SM) and New 
Physics (NP).  As described in Section 2, our approach involves 
the mixing mass difference $\Delta M_{B_s}$ as well.  
The `Update' mentioned in the title refers to two items.  
The first is to apply to our analysis the LHC 
bound for the branching ratio ${\cal B}r_{B_s \to \mu^+\mu^-}$ 
which was recently made public by Ref.~\cite{eps2011}.  
For example, this affects the exclusion region in parameter 
spaces of NP models ({\it cf} Figs.~3,5).
The second update refers to 
a change in one of the inputs in the published article of 
Ref.~\cite{Golowich:2011cx} which modifies our determination of 
$\Delta M_{B_s}$ ({\it cf} Eq. (\ref{me})) and some other  
some of the numerical results displayed in this report.  This 
is also taken into account 
in Version 2 of the arXiv listing of Ref.~\cite{Golowich:2011cx}.

Since this is a written version of a talk given at DPF 2011, I 
avoid theoretical complexities and 
limit the number of equations; these can be accessed 
in Ref.~\cite{Golowich:2011cx}.  However, I will use this 
report to expand on the limited number of references shown in 
my DPF talk by including substantially more here.

Section 2 contains a summary of the current experimental situation 
regarding $B_s$ mixing and ${B_s \to \mu^+\mu^-}$ as well as 
SM predictions for ${\cal B}r_{B_s \to \mu^+\mu^-}$ and 
$\Delta M_{B_s}$.  Most of the underlying theoretical formalism 
and derived relationships in this area are already well known.  
What is more important in comparing theory with current experiment is 
instead {\it numerics}, especially regarding the uncertainties encountered 
in SM predictions. 

In Section 3, we describe how the comparison of 
$\Delta M_{B_s}^{\rm (SM)}$ with $\Delta M_{B_s}^{\rm (Expt)}$ 
yields phenomenological 
bounds on the NP $B_s$-mixing contribution $\Delta M_{B_s}^{\rm (NP)}$.
This can, in turn, 
be used to constrain the magnitude of NP contributions to 
the $B_s \to \mu^+ \mu^-$ decay mode.  
We implement this approach by exploring five possible NP models:  
an extra gauge boson $Z'$, family symmetry, 
R-parity violating supersymmetry, a fourth sequential quark generation 
and flavor-changing Higgs models.

Our Summary appears in Section 4.

\section{Standard Model Analysis and Experimental Review}
The numerical predictions we cite 
for ${\cal B}r_{B_s \to \mu^+\mu^-}^{\rm (SM)}$,  
$\Delta M_{B_s}^{\rm (SM)}$, {\it etc} are based in part 
upon inputs displayed in Table~\ref{tab:inputs}.  
These entries were the ones available when the analysis in 
Ref.~\cite{Golowich:2011cx} was being performed.  
Of course, improvements in such inputs over time will 
modify our results.

\begin{table}[h]
\begin{center}
\caption{Some Input Parameters}
\begin{tabular}{|l|l|}
\hline $\alpha_s (M_Z) = 0.1184 \pm 0.0007$ \cite{Bethke:2009jm} & 
$|V_{ts}| = 0.0403^{+0.0011}_{-0.0007}$ \cite{PDG} \\
\hline $\Delta M_{B_s} = 
(117.0 \pm 0.8) \times 10^{-13}~{\rm GeV}$ \cite{PDG} & 
$\Delta \Gamma_{B_s}/\Gamma_{B_s} = 0.092_{-0.054}^{+0.051}$
\cite{PDG}   \\
\hline ${\bar m}_t ({\bar m}_t) = (163.4 \pm 1.2)~{\rm GeV}$
 \cite{Melnikov:2000qh} & 
$f_{B_s}\sqrt{{\hat B}_{B_s}} = 275 \pm 13$~MeV  \cite{Laiho:2009eu}\\
\hline ${\hat B}_{B_s} = 1.33 \pm 0.06$  \cite{Laiho:2009eu} & 
$f_{B_s} = 0.2388 \pm 0.0095$~GeV \cite{Laiho:2009eu} \\
\colrule
\end{tabular}
\vskip .05in\noindent
\label{tab:inputs}
\end{center}
\end{table}
\noindent In Table~\ref{tab:inputs}, 
the values for the Cabibbo-Kobayashi-Maskawa (CKM) matrix 
elements $|V_{ts}|$ and $|V_{tb}|$ are taken from the global fit 
of Eq.~(11.27) in Ref.~\cite{PDG}.  Another route to $|V_{ts}|$ 
results in the value given in Eq.~(11.13) of 
Ref.~\cite{PDG}, $|V_{ts}| = 0.0387 \pm 0.0021$.  
In this instance, we have adopted the one in Table~\ref{tab:inputs} 
because it has the smaller listed uncertainty.  Finally, 
corresponding to the running mass ${\bar m}_t ({\bar m}_t)$ 
in Table~\ref{tab:inputs} is 
the top quark pole mass $m_t^{\rm (pole)} = 173.1 \pm 1.3$.

\subsection{About $\bm{B_s \to \mu^+\mu^-}$}

Consider first the SM prediction for ${\cal B}r_{B_s \to \mu^+\mu^-}$.  
A direct calculation of this quantity in the SM gives 
\begin{equation}
{\cal B}r_{B_s \to \mu^+\mu^-}^{\rm (SM)} = 
{1 \over 8 \pi^5} \cdot {M_{B_s}  \over \Gamma_{B_s}} 
\cdot \left(G_F^2 M_W^2 m_\mu f_{B_s} |V_{\rm ts}^* V_{\rm tb}| \eta_Y 
Y({\bar x}_t)\right)^2 \left[ 1 - 4 {m_\mu^2 \over M_{B_s}^2} \right]^{1/2}
\ \ , \nonumber 
\end{equation}
where $\eta_Y$ is a QCD factor, $Y({\bar x}_t)$ is an Inami-Lin
function~\cite{Inami:1980fz} and 
${\bar x}_t \equiv {\bar m}_t^2({\bar m}_t)/M_W^2$ where 
${\bar m}_t$ is the running top-quark mass 
in ${\overline{{\rm MS}}}$ renormalization.  
As pointed out in Ref.~\cite{Buras:2003td}, 
inserting the SM expression for 
$\Delta M_{B_s}$ removes the $|V_{\rm ts}^* V_{\rm tb}|^2$ factor, 
thereby reducing the uncertainty in the SM branching ratio expression.
It is this modified expression we use to 
obtain\footnote{For another recent evaluation, 
see Ref.~\cite{Buras:2010mh}.}    
\begin{equation}
{\cal B}r_{B_s \to \mu^+\mu^-}^{\rm (SM)} = 
{3 \over 4 \pi^3} \cdot {\Delta M_{B_s}^{\rm (Expt)}  
\over \Gamma_{B_s}} \cdot {(G_F M_W m_\mu \eta_Y Y )^2 
\over {\hat \eta} {\hat B}_{B_s} S_0({\bar x}_t)}
\left[ 1 - 4 {m_\mu^2 \over M_{B_s}^2} \right]^{1/2}
= \left(3.3 \pm 0.2 \right) \times 10^{-9} \ \ , 
\label{bsmumu}
\end{equation}
where the largest source of uncertainty arises from ${\hat B}_{B_s}$ 
followed by the dependence in $S_0({\bar x}_t)$ 
({\it cf.} discussion preceding Eq.~(\ref{me})) on the $t$-quark mass.


In response to a request from a Session Organizer, my talk 
included a summary of the experimental situation for 
${\cal B}r_{B_s \to \mu^+\mu^-}$.  It has begun to change 
rather substantially.  Below are values taken 
respectively from the Particle Data Group (PDG)~\cite{PDG}, 
the CDF collaboration~\cite{Aaltonen:2011fi} and a compilation (LHC)
of LHCb and CMS data~\cite{eps2011}, all in units of $10^{-9}$: 
\begin{eqnarray}
& & {\cal B}r_{B_s \to \mu^+\mu^-}^{\rm (PDG)} < 47 \times 10^{-9} 
\qquad (\text{CL} \ = 90\%) ~; \qquad 
{\cal B}r_{B_s \to \mu^+\mu^-}^{\rm (CDF)} = 
\left( 18^{+11}_{-9} \right) \times 10^{-9} \nonumber \\
& & {\cal B}r_{B_s \to \mu^+\mu^-}^{\rm (LHC)} < 9 \times 10^{-9} 
\qquad (\text{CL} \ = 90\%)~; \qquad 
{\cal B}r_{B_s \to \mu^+\mu^-}^{\rm (LHC)} < 11 \times 10^{-9} 
\qquad (\text{CL} \ = 95\%) \ \ . \label{data}
\end{eqnarray}
The strongest bounds (from LHC), which have become publically available 
at the 2011 summer conferences, show just how rapidly data for the 
$B_s \to \mu^+\mu^-$ mode is being accumulated.  The `NP Window', 
currently  
\begin{equation}
{\cal B}r_{B_s \to \mu^+\mu^-}^{\rm (LHC)}/
{\cal B}r_{B_s \to \mu^+\mu^-}^{\rm (SM)} \simeq 3.3 \ \ ,
\nonumber
\end{equation}
is expected to close sometime in 2012 if ${B_s \to \mu^+\mu^-}$ 
is not observed by then~\cite{Bettler}.  As such, we are either 
getting near to encountering NP in this mode or testing the 
SM prediction.  Either outcome is eagerly awaited.

\subsection{About $\bm{\Delta M_{B_s}}$}

\noindent The experimental value~\cite{PDG} for $\Delta M_{B_s}$, 
\begin{equation}
\Delta M_{B_s}^{\rm (Expt)} = (117.0 \pm 0.8) \times 10^{-13}~{\rm GeV} 
 \ \ , 
\label{mexpt}
\end{equation} 
is a very accurate one -- the uncertainty amounts to about $0.7$\%.
The next-to-leading (NLO) SM formula is arrived at from an operator product 
expansion of the mixing 
hamiltonian~\cite{{Buras:1990fn},{Urban:1997gw}}.  
The short-distance dependence in 
the Wilson coefficient appears in the scale-insensitive 
combination $\eta_{B_s} 
S_0({\bar x}_t)$, where 
the factor $S_0({\bar x}_t)$ is another Inami-Lin
function~\cite{Inami:1980fz}. Our determination yields 
$S_0({\bar x}_t) = 2.319 \pm 0.028$.  Also, we obtain 
$\eta_{B_s} = 0.5525 \pm 0.0007$ for the NLO QCD factor.  From the 
numerical inputs discussed thus far, we find 
\begin{equation}
\Delta M_{B_s}^{\rm (SM)} = 
{\left( G_{\rm F} M_{\rm W} |V_{\rm ts}^* V_{\rm tb}| 
\right)^2 \over 6 \pi^2} M_{B_s} 
 f^2_{B_s}{\hat B}_{B_s} \eta_{B_s} S_0({\bar x}_t) = 
\left( 125.2^{+13.8}_{-12.7}\right) \times 10^{-13}~{\rm GeV} 
\label{me}
\end{equation} 
which is in accord with the experimental value of 
Eq.~(\ref{mexpt}).


\section{New Physics Analysis}
Let us first obtain a numerical (1$\sigma$) bound 
on the New Physics contribution to 
$\Delta M_{B_s}$.\footnote{The possibility of utilizing 
$\Delta \Gamma_{B_s}$ is not considered here, but is addressed 
in Ref.~\cite{Golowich:2011cx}.}  We then 
use this to constrain couplings in a variety of NP models 
and thereby learn something about the $B_s \to \mu^+\mu^-$ transition.

Accounting for NP as an additive contribution, 
\begin{equation}
\Delta M_{B_s}^{\rm (Expt)} =  \Delta M_{B_s}^{\rm (SM)} + 
\Delta M_{B_s}^{\rm (NP)} \ \ ,   \nonumber 
\end{equation}
we have from Eqs.~(\ref{mexpt}),(\ref{me}), 
\begin{eqnarray}
& & \Delta M_{B_s}^{\rm (NP)} 
= \left(-8.2^{+13.8}_{-12.7}\right) \times 10^{-13}~{\rm GeV} \ \ . 
\nonumber 
\end{eqnarray}
The error in $\Delta M_s^{\rm (expt)}$ has been included, but it  
is so small compared to the theoretical error 
in $\Delta M_s^{\rm (SM)}$ as to be negligible.
The $1\sigma$ range for the NP contribution is thus
\begin{eqnarray}
& & \Delta M_{B_s}^{\rm (NP)} = 
(- 20.9 \to + 5.6)\times 10^{-13}~{\rm GeV} \ \ .
\nonumber 
\end{eqnarray}
To proceed further without ambiguity, we would need to know the 
relative phase between the SM and NP components.  Lacking this, 
we employ the absolute value of the largest possible number,  
\begin{eqnarray}
& & |\Delta M_{B_s}^{\rm (NP)}| \le 
20.9 \times 10^{-13}~{\rm GeV} \ \ ,
\label{bnd}
\end{eqnarray}
to constrain the NP parameters. 

Next is the issue of {\it which} model of NP to adopt.  There 
are, in fact, a number of ways that NP can impact the SM:
\begin{itemize}
\item Extra gauge bosons (LR models, {\it etc})
\item Extra scalars (Multi-Higgs models, {\it etc})
\item Extra fermions (Little Higgs models, {\it etc})
\item Extra dimensions (Universal extra dimensions, {\it etc})
\item Extra global symmetries (Supersymmetry, {\it etc})
\end{itemize}
For an analysis of $D^0$ mixing which includes as many as twenty-one 
NP models, see Ref.~\cite{Golowich:2007ka}.  Here, however, we shall 
content ourselves with a smaller NP menu of five items.
For each, we shall include a few, somewhat informal, introductory 
remarks.

Many NP models have multidimensional parameter spaces whose 
complexity hinders the ability to utilize the constraint of 
Eq.~(\ref{bnd}).  Several strategies come to mind 
for addressing this situation.  Our approach is possibly 
the simplest one -- to employ whatever set of 
reasonable assumptions (and/or physical intuitions) which 
allow us to find paths in parameter 
space which relate $|\Delta M_{B_s}^{\rm (NP)}|$ to 
${\cal B}r_{B_s \to \mu^+\mu^-}$.  However, 
we plan to revisit this issue in future work.


\subsection{Single Extra Vector Boson $\bm{Z'}$}

\begin{figure} [tb]
\centerline{
\includegraphics[width=5cm,angle=0]{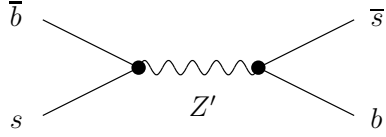}}
\caption{$B_s$ Mixing via $Z'$ exchange}
\label{figure1}
\end{figure}

\noindent We should clarify what is meant by the title of this 
subsection.  There are many $Z'$ models, literally hundreds, 
which contain one or more of these hypothetical gauge bosons.  
What we have in mind is the subset of such models in which 
one $Z'$ has a much lower mass than other NP degrees of 
freedom.  This explains the `single' extra vector boson $Z'$.
In this case, the NP contribution to $B_s$ mixing arises from the 
$Z'$ pole diagram ({\it cf} Fig.~\ref{figure1}).  
Moreover we assume that the $Z'$ has SM couplings to lepton pairs, 
thus leaving us with two unknowns, the $Z'$ mass 
$M_{Z'}$ and the nondiagonal flavor coupling $g_{Z's{\bar b}}$.  

The scaling with $Z'$ parameters then goes as 
\begin{equation}
\Delta M_{B_s}^{\rm (Z')} \ \propto\ 
{g_{Z's{\bar b}}^2 \over M_{Z'}^2} \qquad \text{and} \qquad 
{\cal B}r_{B_s \to \mu^+\mu^-}^{\rm ({Z'})} \ \propto \ 
{g_{Z's{\bar b}}^2 \over M_{Z'}^2}\cdot {M_Z^2 \over M_{Z'}^2} \ \ , 
\nonumber 
\end{equation} 
leading to 
\begin{equation}
{\cal B}r_{B_s \to \mu^+\mu^-}^{\rm ({Z'})} \le 
0.3 \times 10^{-9}~\cdot \left( {1~{\rm TeV} \over M_{Z'}}\right)^2 \ \ .
\nonumber 
\end{equation} 
This value will lie below the corresponding 
SM prediction (${\cal B}r_{B_s \to \mu^+\mu^-}^{\rm (SM)}  = 
3.3 \times 10^{-9}$) even for a $Z'$ mass as light as 
$M_{Z'} > 0.3$~TeV.  So, this class of models will not 
dominate the SM result.



\subsection{Family (`Horizontal') Symmetries} 

The motivation for Family Symmetries (FS) is to obtain an 
understanding (still needed!) of fermion masses and fermion 
mixing matrices.
This subject has a relatively long history, first having been 
actively explored in the late 1970's.  However, it remains an 
area of theoretical interest up to recent times ({\it e.g.} 
see Ref.~\cite{Lalak:2010bk} where FCNC and CPV patterns are 
studied in the context of Family Symmetries).

Here is a brief overview, with details left for 
Ref.~\cite{Golowich:2011cx}.    
The gauge sector in the Standard Model has a large global symmetry which
is broken by the Higgs interaction~\cite{Sher:1988mj}.  
By enlarging the Higgs sector,
some subgroup of this symmetry can be imposed on the full SM lagrangian
and the symmetry can be broken spontaneously.  This family symmetry can be
global~\cite{fcnch} as well as gauged~\cite{Maehara:1977nq}.
If the new gauge couplings are very weak
or the gauge boson masses are large, the difference between a gauged or
global symmetry is rather difficult to distinguish 
in practice~\cite{begmemor}.  In 
general there would be FCNC effects from both the gauge and scalar
sectors.  GHPPY analyze the gauge contribution.  Consider the family
gauge symmetry group $SU(3)_G$ acting on the three left-handed 
families.  Spontaneous symmetry breaking will render all the gauge bosons
massive.  If the SU(3) is broken first to SU(2) before being
completely broken, we may have an effective `low' energy 
symmetry $SU(2)_G$.  This means that the gauge bosons 
${\bf G} \equiv \left\{ G_i \right\} \ 
(i = 1,\dots ,3)$ have masses $m_i$ which are much lighter 
than those of the 
$\left\{ G_k \right\} \ (k = 4,\dots ,8)$.

There is a history of applications of family symmetry 
in which the number of unknowns becomes reduced to 
manageable proportions.  The whole story is rather involved, 
but the following gives an impression of some steps.  
As regards the gauge boson masses $\{ m_i \}$, 
in a simple scheme of symmetry breaking~\cite{Monich:1980rr} 
one obtains $m_1 = m_3$.  There are in the family symmetry also 
four mixing matrices $U_d, U_u, U_\ell$, $U_\nu$, unknown except for 
the constraints 
\begin{equation}
U_u^\dagger U_d = V_{\rm CKM} \qquad \text{and} \qquad  
U_\nu^\dagger U_\ell \ = V_{\rm MNSP} \ \ ,
\nonumber
\end{equation}
where $V_{\rm CKM}$ and $V_{\rm MNSP}$ are respectively 
the well-known 
Cabibbo-Kobayashi-Maskawa and Maki-Nakagawa-Sakata-Pontcorvo 
mixing matrices for quarks and leptons. 
Through reasoning given in Refs.~\cite{q7},~\cite{Harrison:2002er}, 
it is somewhat possible to tame the zoo of unknown mixing parameters.

The upshot of all this is 
\begin{equation}
\Delta M_{B_s}^{\rm (FS)} \ \propto\ {f^2 \over m_1^2} \qquad 
\text{and} \qquad 
{\cal B}r_{B_s \to \mu^+\mu^-}^{\rm ({FS})} \ \propto \ 
{f^4 \over m_1^4} \ \ , 
\nonumber 
\end{equation} 
where $f$ sets the scale of the interactions between the gauge-bosons 
and fermions.  The above relations then yield the result 
\begin{equation}
{\cal B}r_{B_s \to \mu^+\mu^-}^{\rm ({FS})} \le 
0.9 \times 10^{-12} \ \ ,
\nonumber 
\end{equation} 
which is tiny compared to the SM prediction.


\subsection{R-parity Violating Supersymmetry}

\begin{figure} [tb]
\centerline{
\includegraphics[width=6cm,angle=0]{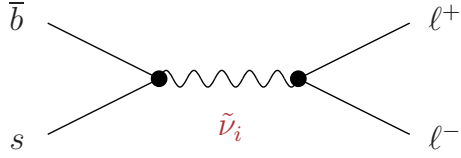}}
\caption{$B_s \to \ell^+\ell^-$ via sneutrino exchange}
\label{figure2}
\end{figure}

One of the models of New Physics that has a rich flavor phenomenology 
is R-parity violating (RPV) SUSY. R-parity distinguishes between 
particle and sparticle as
\begin{equation}
R_P = (-)^{3(B-L) + 2S} = \left\{ 
\begin{array}{ll}
+1 & \text{(particle)} \\
-1 & \text{(sparticle)} 
\end{array}
\right.
\nonumber
\end{equation}
If R-parity is conserved and the initial state consists solely 
of ordinary matter, then intermediate and final states can 
contain only even numbers of sparticles.  In this subsection, 
we lift this restriction by allowing the presence in the
superpotential of terms 
\begin{equation} \nonumber
{\cal W} _{\not R}= \frac{1}{2} \lambda_{ijk} L_i L_j E^c_k +
 \lambda_{ijk}^\prime L_i Q_j D^c_k + 
 \frac{1}{2} \lambda_{ijk}^{\prime\prime} U_i^c D^c_j D^c_k 
\end{equation}
where $i=1,2,3$ are generation labels.
We impose 
baryon number symmetry by setting $\lambda^{\prime\prime}$ to zero. 
Also, we will assume CP-conservation, so all couplings $\lambda_{ijk}$ and 
$\lambda_{ijk}^\prime$ become real-valued. 
The set of $\{\lambda_{ijk}\}$ occur for couplings to leptons whereas 
only the $\{ \lambda_{ijk}^\prime\}$ occur in $B_s$ mixing.
For example, the 
lagrangian describing RPV SUSY contributions to $B_s$ mixing can be 
written as
\begin{equation}
{\cal L}_{\not R}=-\lambda_{i32}^\prime \widetilde\nu_{i_R} 
\overline b_R s_L - 
\lambda_{i23}^\prime \widetilde\nu_{i_R} \overline s_R b_L + h.c.\ \ , 
\nonumber 
\end{equation}
The crucial difference between studies of 
RPV SUSY contributions to phenomenology of the up-quark 
and down-type quark sectors is the possibility of tree-level 
diagrams contributing to 
$B_s$-mixing\footnote{We assume that there is no 
strong hierarchy between the RPV SUSY couplings 
that favors possible box diagrams.} and $B_s \to \ell^+\ell^-$ 
decays~\cite{Kundu:2004cv,Kao:2009fg,Dreiner:2006gu,Saha:2002kt}.
Then a process like the one in Fig.~\ref{figure2} dominates. 
Note how the initial and final states consist of 
ordinary matter whereas the intermediate state has a single 
sneutrino -- this is clearly $B_s$ mixing due to RPV dynamics.


\vspace{0.2cm}

\begin{figure} [tb]
\centerline{
\includegraphics[width=9cm,angle=0]{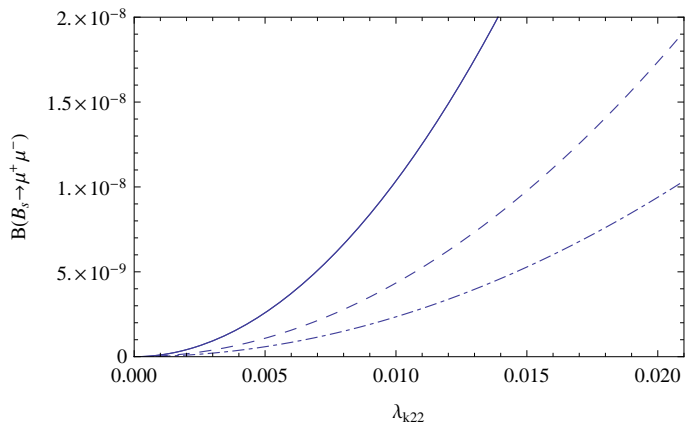}}
\caption{Branching ratio of 
${\cal B}r_{B_s^0 \to \mu^+\mu^-}$ as a function
of RPV leptonic coupling $\lambda_{k22}$ and 
sneutrino mass $M_{\tilde\nu_i} = 100$~GeV,
$150$~GeV, and $200$~GeV (solid, dashed, and dash-dotted lines).}
\label{figure3}
\end{figure}

In RPV-SUSY, the underlying transition for $B_s \to \mu^+\mu^-$ 
is $s + {\bar b} \to \mu^+ + \mu^-$ via tree-level $u$-squark or 
sneutrino exchange.
In order to relate the rare decay to the 
mass difference contribution from RPV SUSY 
$\Delta M_{\rm B_s}^{\rm ({\not R})}$, we need to assume 
that the up-squark contribution is 
negligible. This can be achieved in models where 
sneutrinos are much lighter than the 
up-type squarks, which are phenomenologically viable. 
Then it can be shown that the dependence of 
${\cal B}r^{\rm ({\not R})}_{B_s\to\mu^+\mu^-}$ on the RPV 
parameters becomes 
\begin{equation}
\nonumber
{\cal B}r^{\rm ({\not R})}_{B_s\to\mu^+\mu^-} \ \propto \ 
\left(\frac{\lambda_{i22}\lambda_{i32}^\prime}
{M_{\tilde\nu_i}^2}\right)^2 \ \ .
\end{equation}
Upon inserting the information from 
$\Delta M_{\rm B_s}^{\rm ({\not R})}$ and assuming 
$\lambda_{k23}=\lambda_{k32}$, 
it is possible to plot the dependence of ${\cal B}r_{B_s\to\mu^+\mu^-}$ on 
$\lambda_{k22}$ for different values of $M_{\tilde\nu_i}$, 
which we present in Fig.~\ref{figure3}.  



\subsection{Fourth Quark Generation} 
One of the simplest extensions of the Standard Model involves 
addition of the sequential fourth generation (often denoted 
as SM4) of chiral quarks, $t'$ and 
$b'$~\cite{Holdom:2009rf,Buras:2010pi,Hou:2010mm}.  
The addition of a 
sequential fourth generation of quarks 
leads to a 4$\times$4 CKM quark mixing 
matrix~\cite{Chanowitz:2009mz}. This implies 
that the parameterization 
of this matrix requires six real parameters and three phases. 
Besides providing 
new sources of CP-violation, the two additional phases 
can affect ${\cal B}r_{B_s \to \mu^+\mu^-}$ due to interference 
effects~\cite{Bobrowski:2009ng}. 

\begin{figure} [b]
\centerline{
\includegraphics[width=9cm,angle=0]{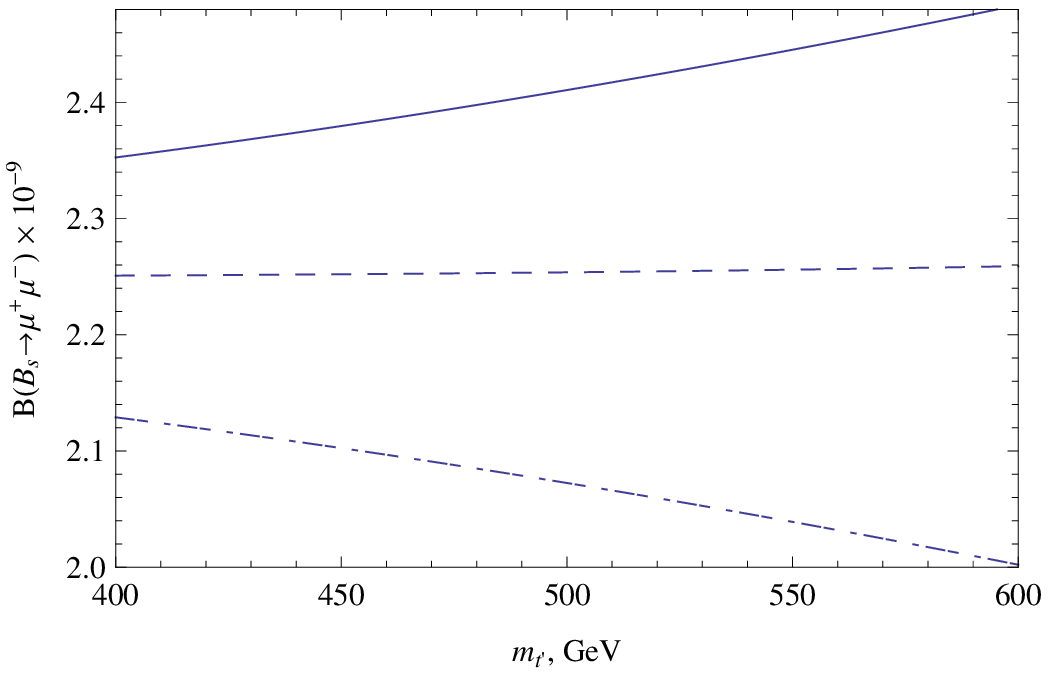}
\includegraphics[width=9cm,angle=0]{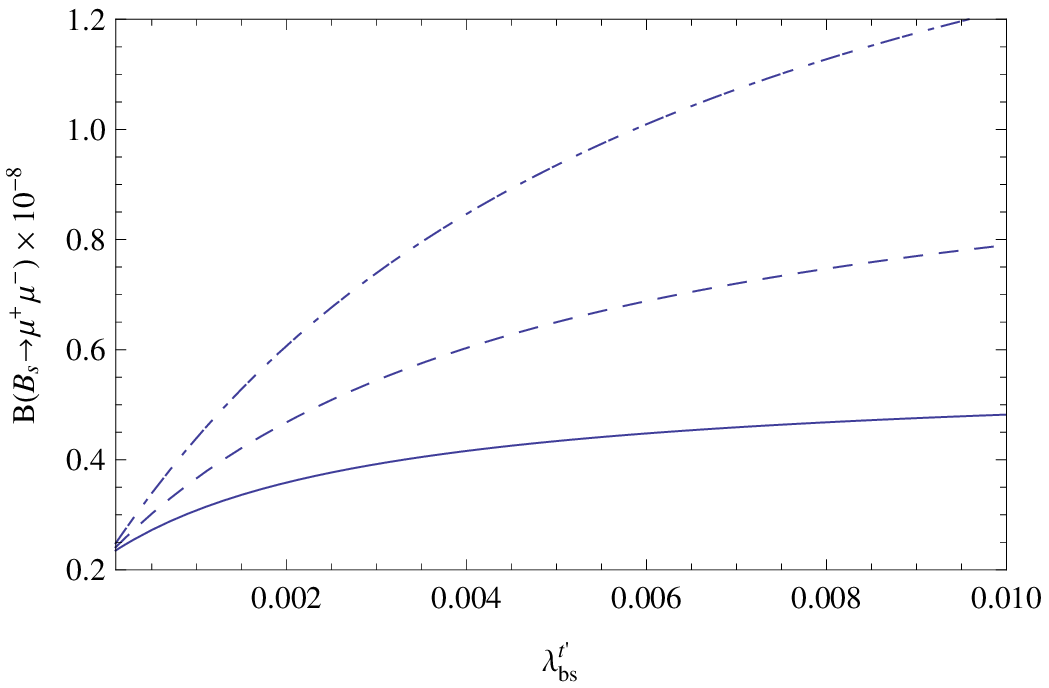}}
\caption{Left: branching ratio of 
${\cal B}r_{B_s^0 \to \mu^+\mu^-}$ as a function
of the top-prime mass $m_{t'}$ for different values of the phase 
$\phi_{t's}=0,\pi/2,\pi$ 
(solid, dashed, dash-dotted lines) and 
$\lambda_{bs}^{t'}=|V_{t's} V_{t'b}^*|\simeq 10^{-4}$~\cite{Alok:2010zj} 
(see also~\cite{Nandi:2010zx}).
Right: branching ratio of ${\cal B}r_{B_s^0 \to \mu^+\mu^-}$ as a function
of the CKM parameter combination $\lambda_{bs}^{t'}$ with 
$\phi_{t's}=0$ and different 
values of $m_{t'}=400$~GeV (solid), $500$~GeV (dashed), 
and $600$~GeV (dash-dotted).}
\label{figure4}
\end{figure}

There are several existing constraints on the parameters 
related to the fourth generation of 
quarks, including direct searches, CKM unitarity tests, 
and fitting precision electroweak data 
(S and T parameters)~
\cite{Novikov:1994zg,Novikov:2001md,Kribs:2007nz,Erler:2010sk}.  
The latter strongly constrains the masses of the new 
quarks.  Finally, for the sake of completeness we take note of 
a data input which became available subsequent to our 
DPF 2011 talk -- that null results in Higgs searches 
by the LHC detectors now place the scenario of a sequential 
fourth quark generation `in deep trouble'~\cite{lp2011}.

The relationship between $\Delta M_{B_s}$ and 
${\cal B}r_{B_s \to \mu^+\mu^-}^{\rm (SM4)}$ has been previously 
studied in detail in Ref.~\cite{Soni:2010xh}. 
We shall update their result.  In SM4, the branching ratio for 
$B_s \to \mu^+\mu^-$
can be related to the experimentally-measured\footnote{We shall 
use $\Delta M_{B_s}^{\rm (Expt)}$, as the 
separation of NP and SM contributions used 
in the rest of this paper, {\it viz} $\Delta M_{B_s} = 
\Delta M_{B_s}^{\rm (SM)} + \Delta M_{B_s}^{\rm (SM4)}$, is not possible
here due to loops with both $t'$ and $t$, $c$, or $u$ quarks.}  
$\Delta M_{B_s}^{\rm (Expt)}$ as~\cite{Soni:2010xh}
\begin{equation}
{\cal B}r_{B_s \to \mu^+\mu^-}^{\rm (SM4)} 
\ \propto \ \Delta M_{B_s}^{\rm (Expt)}
\frac{\left|C_{10}^{tot}\right|^2}{\left|\Delta^\prime\right|} \ \ ,
\nonumber 
\end{equation} 
where the parameter $\Delta^\prime$ is a $B_s$-mixing 
loop parameter~\cite{Soni:2010xh},
\begin{equation}
\Delta^\prime = \eta_t S_0(x_t) + \eta_{t'} R_{t't}^2 S_0(x_{t^\prime}) +
2 \eta_{t^\prime}  R_{t't} S_0(x_t,x_{t^\prime}) \ \ ,
\nonumber 
\end{equation} 
$R_{t't} \equiv V_{t's} V_{t'b}^*/V_{ts} V_{tb}^*$ and 
the definition of the function $S_0(x_t,x_{t^\prime})$ can be 
found in Ref.~\cite{Soni:2010xh}. 
The Wilson coefficient $C_{10}^{tot}$ is defined as
\begin{equation}
C_{10}^{tot}(\mu) = C_{10} (\mu) + R_{t't} C_{10}^{t'}(\mu)
\nonumber 
\end{equation} 
with $C_{10}^{t'}$ obtained by substituting $m_{t'}$ into the 
SM expression for $C_{10}$~\cite{Buras:1994dj}.
Our results can be found in Fig.~\ref{figure4}. 
Referring to 
the recent LHC bound of Eq.~(\ref{data}) on 
${\cal B}r_{B_s\to\mu^+\mu^-}$, the exclusion regions 
on the ${B_s\to\mu^+\mu^-}$ branching ratio in Fig.~4 
lies above $1.\times 10^{-8}$.
As one can see, the resulting branching ratios are
for the most part lower than the LHC experimental bound.  
However, with values of the CKM4 matrix element 
$\lambda_{bs}^{t'}\equiv |V_{t's} V_{t'b}^*|$ of about $0.01$, 
disfavored by~\cite{Alok:2010zj}, 
but still favored by~\cite{Nandi:2010zx}, 
${\cal B}r_{B_s\to\mu^+\mu^-}^{\rm (SM4)}$ 
can still exceed the current bound. 

\subsection{Flavor-changing Neutral Current Higgs Bosons}
Many extensions of the Standard Model contain multiple scalar 
doublets, which increase the
possibility of FCNC mediated by flavor non-diagonal interactions of
neutral components. While
many ideas exist on how to suppress those interactions 
(see, {\it e.g.}~Refs.~\cite{Hall:1993ca,Cheng:1987rs,Pich:2009sp}), 
the ultimate test of those ideas would involve direct 
observation of scalar-mediated FCNC.  The interest in 
multi-Higgs structures has remained fairly constant over the 
years and continues unabated to this day ({\it e.g.} see 
Ref.~\cite{Buras:2010mh}).

A generic interaction hamiltonian of this type is 
\begin{eqnarray}
\nonumber
& & {\cal H}^{H} = \frac{\phi^0}{\sqrt{2}} \left[
\lambda_{23}^{D\dagger}  
\ \overline b_R s_L   \ + 
 \lambda_{32}^D  \ \overline b_L s_R   +
\lambda_{22}^E  \ \overline \mu_L \mu_R \right]
-  i \frac{a^0}{\sqrt{2}} \left[ 
\lambda_{23}^{D\dagger} \ 
\overline b_R s_L   \ + 
i \lambda_{32}^D \ \overline b_L s_R  + \
i \lambda_{22}^E \ \overline \mu_L \mu_R  \right] 
+ \dots  \ + \mbox{H.c.}\ \ ,
\end{eqnarray}
where $\phi^0$ and $a^0$ represent the lightest scalar 
and pseudoscalar states respectively and 
ellipses stand for the terms containing still heavier states 
whose contributions to $\Delta M_{B_s}$ and 
${\cal B}r_{B_s \to \mu^+\mu^-}$ will be suppressed.  
We take all couplings $\lambda_{ij}^D$ and $\lambda_{ij}^E$ 
to be real-valued.  The superscripts $D$ and $E$ on these refer 
respectively to couplings of $d$-type quarks and of charged leptons.
To proceed, we need to distinguish two cases: the 
lightest FCNC Higgs particle as a scalar ($\phi^0$) or 
pseudoscalar ($a^0$).

\begin{figure} [b]
\centerline{
\includegraphics[width=9cm,angle=0]{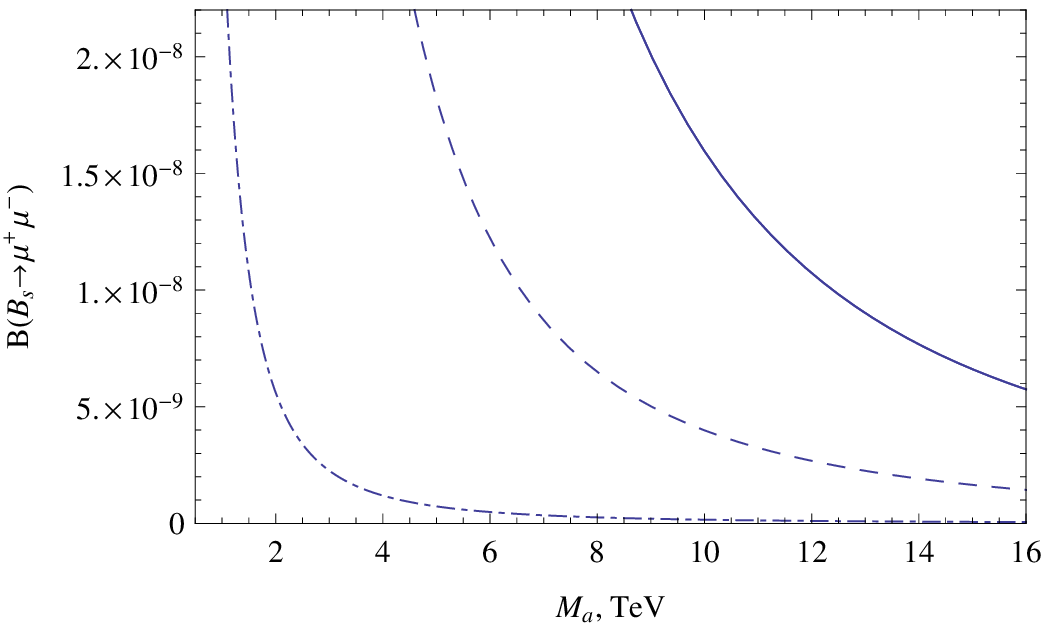}
\includegraphics[width=9cm,angle=0]{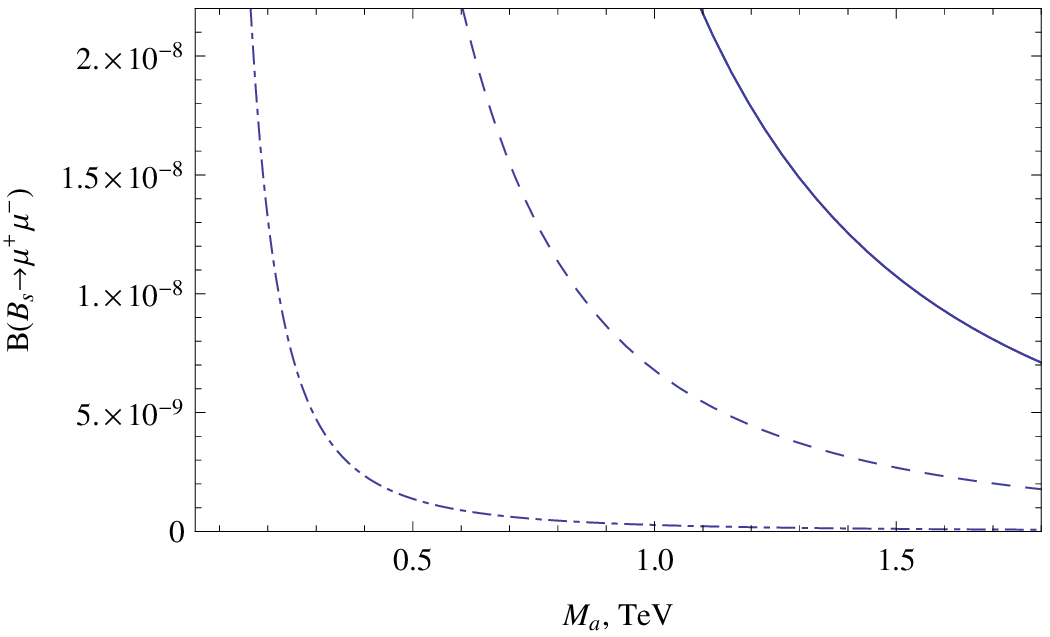}}
\caption{Branching ratio of 
${\cal B}r_{B_s^0 \to \mu^+\mu^-}$ as a function
of pseudoscalar Higgs mass $M_a$. Left: $\lambda_{22}^E=1,0.5,0.1$ 
(solid, dashed, dash-dotted lines).
Right: $\lambda_{22}^E=0.1,0.05,0.01$ 
(solid, dashed, dash-dotted lines).}
\label{figure5}
\end{figure}

\subsubsection{Light scalar FCNC Higgs}

The case of relatively light scalar Higgs state is quite common, 
arising most often in Type-III two-Higgs doublet models (models without
natural flavor 
conservation)~\cite{Barger:1989fj,Atwood:1996vj,Blechman:2010cs}. 
Although the FCNC Higgs model 
does contribute to $\Delta M_{{\rm B}_s}$, it does {\it not} 
contribute to $B_s\to \mu^+\mu^-$ at tree level. 
Any nonzero contribution to
$B_s\to \mu^+\mu^-$ decay must be produced at one-loop
level~\cite{Diaz:2004mk}.

\subsubsection{Light pseudoscalar FCNC Higgs}

The case of a lightest pseudoscalar Higgs state 
can occur in the non-minimal supersymmetric
standard model (NMSSM)~\cite{Nilles:1982dy,Ellis:1988er,
Ellwanger:1996gw,Hiller:2004ii} 
or related models~\cite{Dobrescu:1999gv}. In NMSSM, 
a singlet pseudoscalar is introduced to dynamically solve 
the $\mu$ problem. The resulting pseudoscalar 
can have a mass as light as tens of GeV.  
This does not mean, however, that it necessarily gives the 
dominant contribution to both $B_s$ 
mixing and the $B_s \to \mu^+\mu^-$ decay rate since there can be loop 
contributions from other Higgs states. However, here 
we work in the region of the parameter space where it does.
Calculation reveals 
\begin{eqnarray}
& & \Delta M_{{\rm B}_s}^{\rm (a)} \ \propto \ 
\left({\lambda_{32}^D \over M_a} \right)^2 \qquad \text{and} \qquad 
{\cal B}r_{B_s \to \ell^+\ell^-}^{\rm (a)} \ \propto \ 
\left({\lambda_{32}^D ~\lambda_{22}^E \over M_a^2} \right)^2 
\nonumber
\end{eqnarray}
In ${\cal B}r_{B_s \to \ell^+\ell^-}^{\rm (a)}$, the dependence on 
$\lambda_{32}^D$ can be eliminated by using 
$\Delta M_{{\rm B}_s}^{\rm (a)}$.   The unknown factors then enter 
in the combination $\lambda_{22}^E/M_a$.
In Fig.~\ref{figure5}, we plot the dependence on 
$M_a$ for different values of $\lambda_{22}^E$.

\section{Summary} 
Our talk consisted of two main parts, the (updated) SM evaluations 
of Ref.~\cite{Golowich:2011cx} and the issue of NP contributions 
to $B_s \to \mu^+\mu^-$.  We discuss each of these in turn.

\vspace{0.3cm}

\subsection{Update of SM Evaluations}

As regards $B_s$ mixing, we have 
\begin{eqnarray}
\Delta M_{B_s}^{\rm (Expt)} = (117.0 \pm 0.8) \times 10^{-13}~{\rm GeV} 
\qquad \text{and} \qquad 
\Delta M_{B_s}^{\rm (SM)} = 
\left( 125.2^{+13.8}_{-12.7}\right) \times 10^{-13}~{\rm GeV} 
 \ \ .
\nonumber 
\end{eqnarray}
The uncertainty in the SM result is seen to be about $16$ times larger 
than that in the experimental listing.  It arises 
mainly from the factors $f^2_{B_s}{\hat B}_{B_s}$ and 
$|V_{\rm ts}|^2$ in Eq.~(\ref{me}) and is 
roughly equally shared between them.  This large theoretical 
uncertainy in $\Delta M_{B_s}^{\rm (SM)}$ 
hinders the study of additive NP contributions.  

\vspace{0.15cm}

As regards the $B_s \to \mu^+\mu^-$ branching ratio, upon  
using $\Delta M_{B_s}^{\rm (Expt)}$ as input 
({\it cf} Eq.~(\ref{bsmumu})), we obtain 
\begin{eqnarray}
{\cal B}r_{B_s \to \mu^+\mu^-}^{\rm (SM)} = 
\left(3.3 \pm 0.2 \right) \times 10^{-9} \ \ . \nonumber
\end{eqnarray}
We find the main uncertainty to be from 
${\hat B}_{B_s}$ and roughly half as much from the 
implicit $t$-quark mass dependence in $S_0({\bar x}_t)$.  

\subsection{NP Contributions to $\bm{B_s \to \mu^+\mu^-}$} 
The GHPPY approach to $B_s \to \mu^+\mu^-$ is to use 
\begin{eqnarray}
& & \Delta M_{B_s}^{\rm (Expt)} = \Delta M_{B_s}^{\rm (SM)} + 
\Delta M_{B_s}^{\rm (NP)} \nonumber 
\end{eqnarray}
to bound $|\Delta M_{B_s}^{\rm (NP)}|$ and thus to constrain 
NP parameters.  In this talk, I described the results of 
studying five NP models, with a sixth in preparaion.  For two 
of them (the cases of a single $Z^{'}$ and of family symmetry), 
we conclude that the contribution to $B_s \to \mu^+\mu^-$ 
lies below the SM prediction.  For the others, Figs.~3-5 depict 
the regions of exclusion in their respective parameter spaces.  
We repeat our earlier comment that the exclusion regions in 
earlier versions~\cite{Golowich:2011cx} of our Figures 
are based on the PDG listing 
\begin{eqnarray}
& & {\cal B}r_{B_s \to \mu^+\mu^-}^{\rm (PDG)} < 47 \times 10^{-9} 
\qquad (\text{CL} \ = 90\%) \nonumber 
\end{eqnarray}
and have been updated here to reflect the combined LHCb and CMS bounds, 
\begin{eqnarray}
& & {\cal B}r_{B_s \to \mu^+\mu^-}^{\rm (LHC)} < \ 
\left\{
\begin{array}{ll}
9 \times 10^{-9} & \qquad (\text{CL} \ = 90\%) \\
11 \times 10^{-9} & \qquad (\text{CL} \ = 95\%) \ \ .\\
\end{array}
\right.
\nonumber 
\end{eqnarray}

As a final comment, we point out that the size of 
the SM contribution to $D^0 \to \mu^+\mu^-$~\cite{Golowich:2009ii} 
is quite feeble compared to that in  $B_s \to \mu^+\mu^-$.  It 
would appear that the $D^0 \to \mu^+\mu^-$ decay mode could 
well be a fruitful arena to search for New Physics!

\begin{acknowledgments}
The work of E.G. was supported in part by the U.S.\ National Science
Foundation under Grant PHY--0555304, J.H. was supported by the U.S.
Department of Energy under Contract DE-AC02-76SF00515, 
S.P. was supported by the U.S.\ Department of 
Energy under Contract DE-FG02-04ER41291 and 
A.A.P.~was supported in part by the U.S.\ National Science Foundation under
CAREER Award PHY--0547794, and by the U.S.\ Department of Energy 
under Contract DE-FG02-96ER41005.  We thank Diego Tonelli for his 
helpful communication.
\end{acknowledgments}

\bigskip 

\end{document}